      [1999/12/01 v1.4c Il Nuovo Cimento]
\documentclass[varenna]{cimento}
\usepackage{graphicx,citesort}
\title{Quantum phase transitions in 
one-dimensional\\ electron-phonon systems}
\author{Holger Fehske}
\institute{Institut f\"ur Physik, 
Ernst-Moritz-Arndt-Universit\"at Greifswald,
17487 Greifswald, Germany}
\author{Eric Jeckelmann}
\institute{Institut f\"ur Theoretische Physik, Universit\"at Hannover, 
30167 Hannover, Germany }

\shortauthor{H. Fehske \atque E. Jeckelmann}
\begin{document}
\maketitle
\section{Introduction}
Low dimensional strongly coupled electron-phonon systems like 
MX-chains, ferroelectric perovskites, conjugated polymers, or
organic charge transfer salts exhibit a remarkable wide range of strengths 
of competing forces and, as a result, physical properties~\cite{TNYS90}. 

\begin{figure}[t]\centerline{
\includegraphics[width=.45\linewidth,clip]
{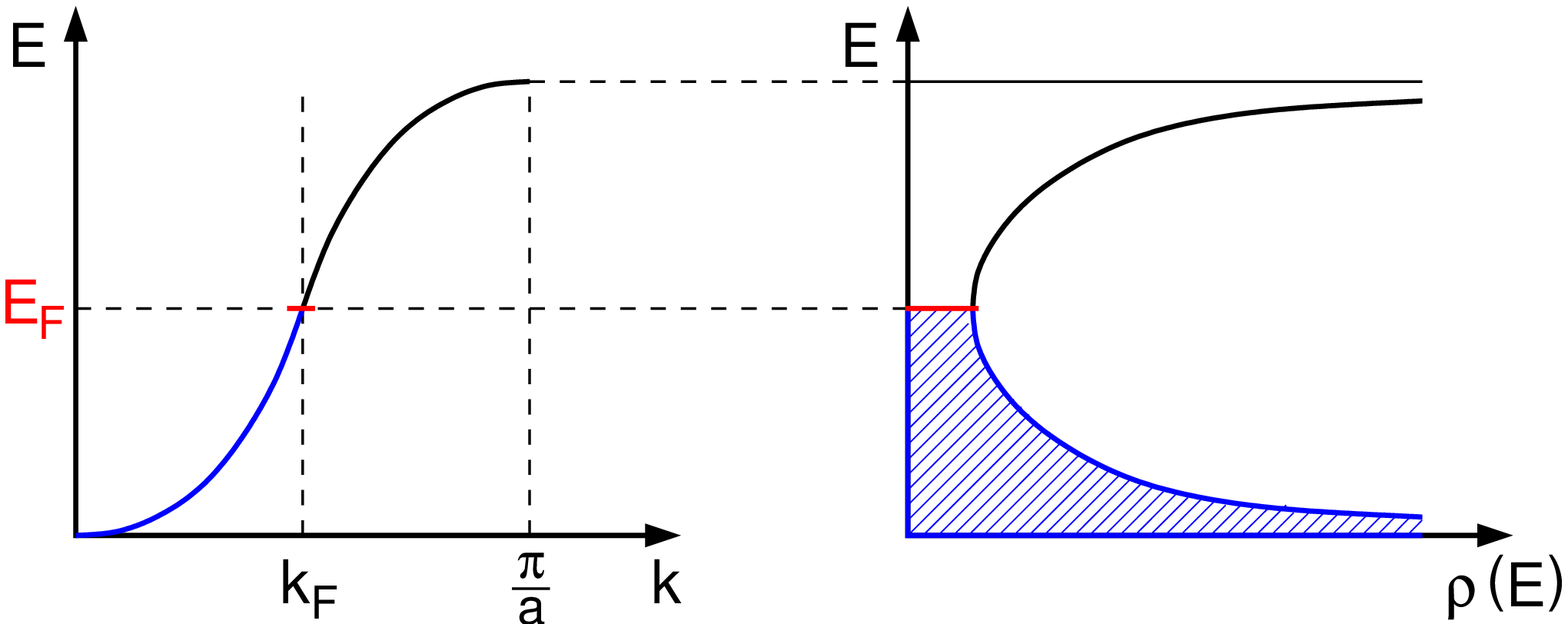}\hspace*{.5cm}
\includegraphics[width=.45\linewidth,clip]
{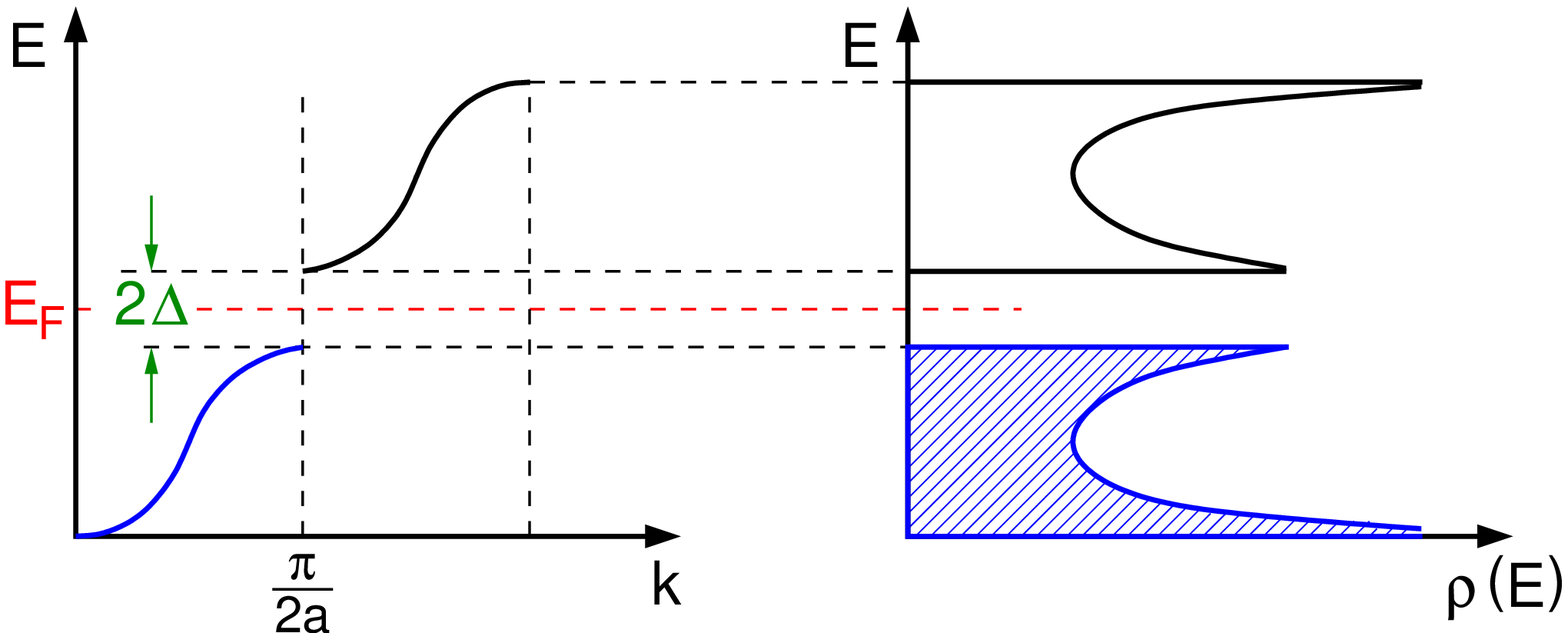}}
\caption{Peierls scenario: A gap $2\Delta$ opens in the electronic 
band structure $E(k)$ [density of states $\rho(E)$] of 
an 1D metal if, as a result of the 
EP coupling, a static lattice distortion occurs, implicating
a new lattice period $2a$ in real space.}
\label{fig_peierlsscenario}
\end{figure}

Most notably quasi one-dimensional (1D) materials are  
very susceptible to structural distortions driven by the 
electron-phonon (EP) interaction. Probably the most famous one is 
the Peierls instability~\cite{Pe55} of 1D metals: 
As the temperature is lowered the system creates a 
periodic variation in the carrier density by shifting
the ions from their symmetric positions. 
For the half-filled band case this so-called
charge density wave (CDW) is commensurate with the lattice,
the unit cell doubles, and the system possesses a spontaneous
broken-symmetry ground state. Since a static dimerisation 
of the lattice opens a gap at the Fermi surface the metal 
gives way to a Peierls insulator (PI) [see Fig.~\ref{fig_peierlsscenario}]. 

The on-site Coulomb interaction, on the other hand,  tends to immobilise 
the charge carriers as well by establishing a Mott 
insulating ground state. The Mott insulator (MI) exhibits strong
spin density wave (SDW) correlations but has continuous symmetry 
and therefore shows no long-range order in 1D. 
Then, of course, the question arises, whether the PI and MI phases 
are separated by one (or more than one) quantum critical point(s) 
at $T=0$, and if so, how the cross-over is 
modified by quantum phonon effects.     

\begin{figure}[b]
\includegraphics[width=.29\linewidth,clip]
{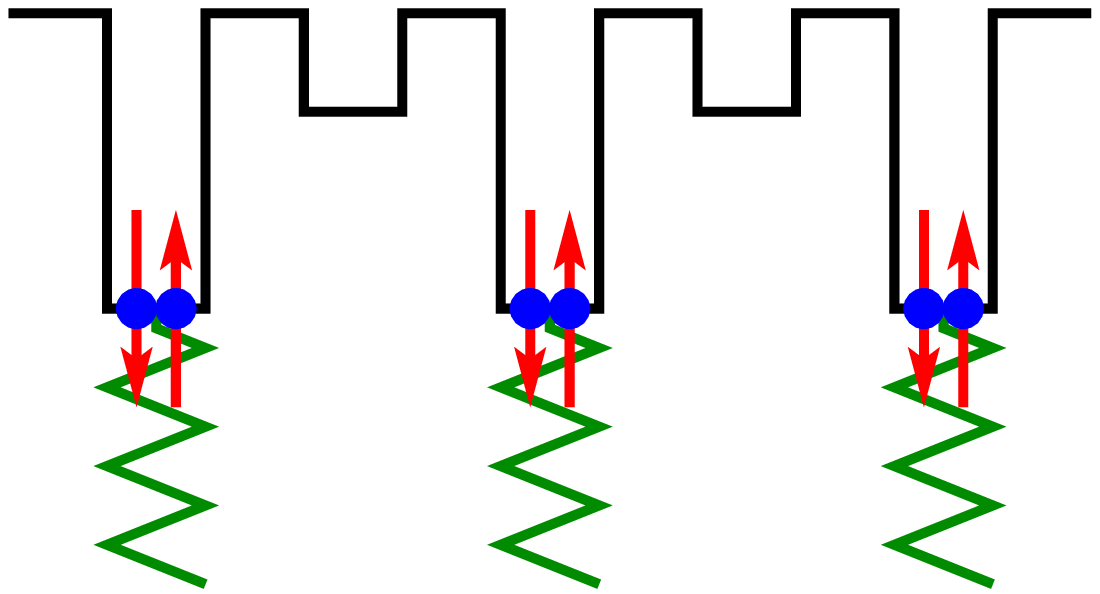}  
\includegraphics[width=.4\linewidth,clip]
      {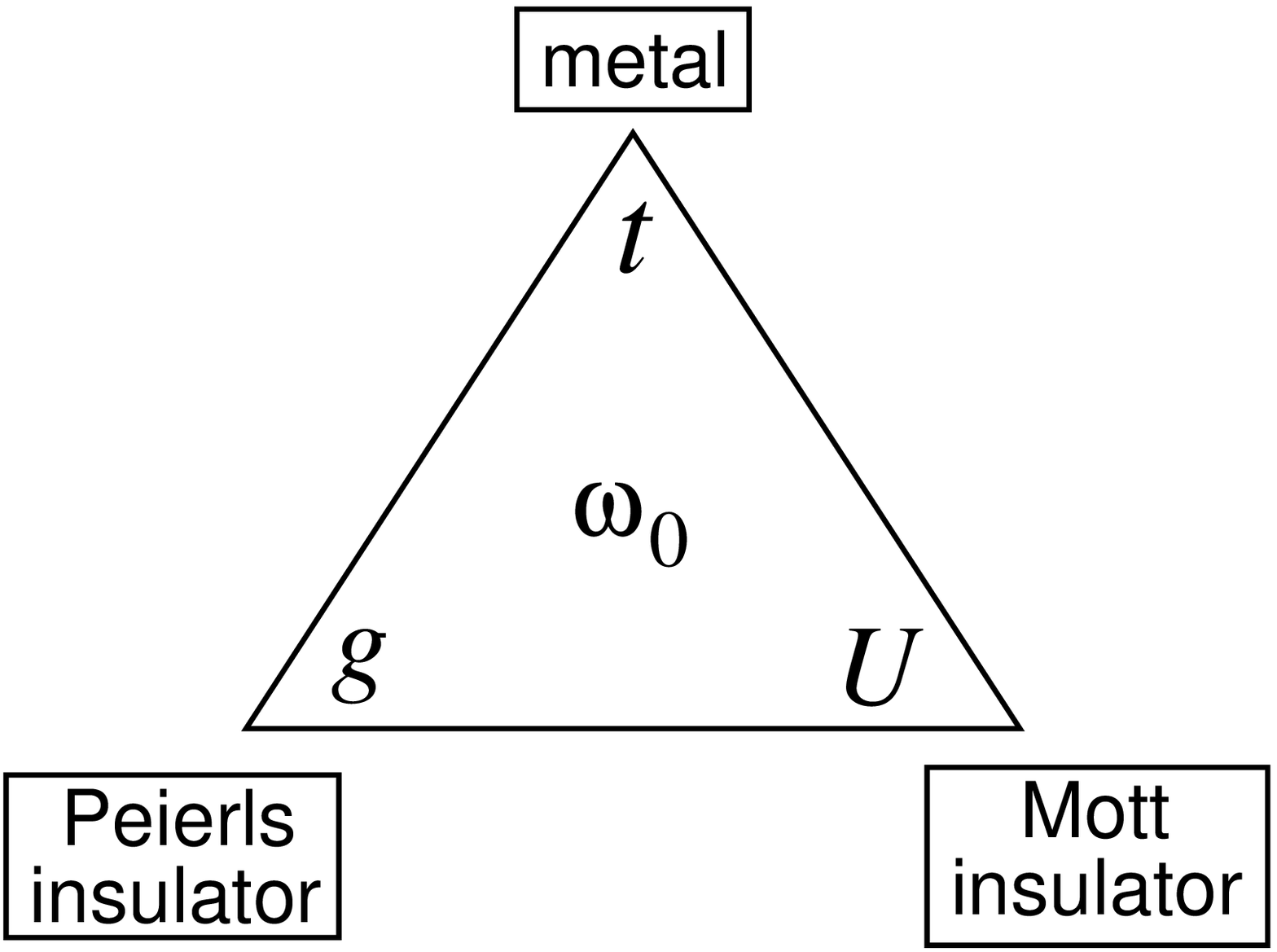}
\includegraphics[width=.29\linewidth,clip]
{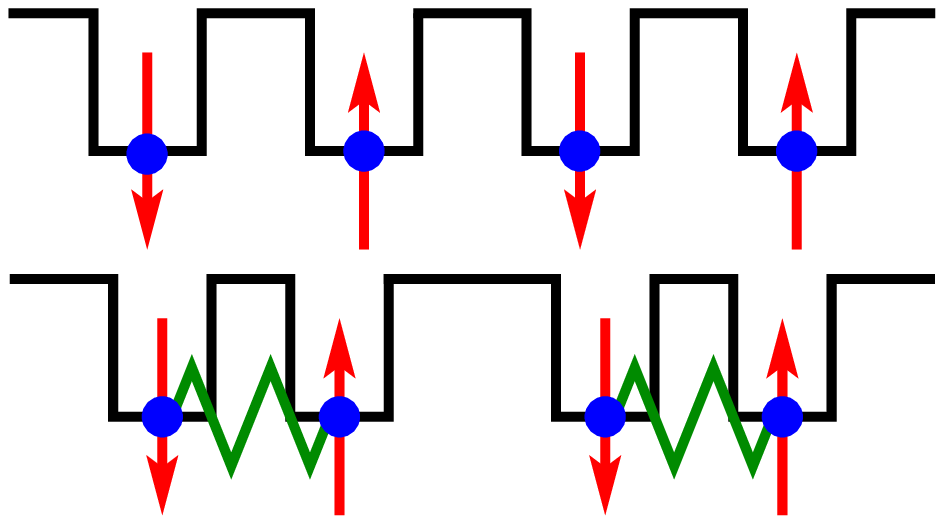}
\caption{Schematic phase diagram of the 1D Holstein Hubbard model.
At half-filling, Peierls (left) or Mott (right) 
insulating phases may be favoured over the metallic state.
In the case of localised electrons interacting via antiferromagnetic
exchange and magneto-elastic couplings even a spin-Peierls distorted state
can emerge (right, lower panel).}
\label{fig_hhm_phys}
\end{figure}

The challenge of understanding such quantum phase transitions
has stimulated intense work on generic microscopic models of 
interacting electrons and phonons like the Holstein Hubbard model (HHM): 
\begin{equation}
 H =  - t
      \sum\limits_{\langle i, j\rangle\sigma}\! c_{i\sigma}^\dagger 
      c_{j\sigma}^{} 
      - g \omega_0
      \sum\limits_{i\sigma} (b_i^\dagger + b_i^{}) n_{i\sigma}^{}
      + \omega_0 \sum\limits_i b_i^\dagger b_i^{}  
      + U
      \sum\limits_i n_{i\uparrow} n_{i\downarrow}\,.
\label{hhm}
    \end{equation}
Here $c^{\dagger}_{i\sigma}$ ($c^{}_{i\sigma}$) denote
fermionic creation (annihilation) operators of electrons with 
spin $\sigma=\uparrow,\,\downarrow$ on a 1D lattice with $N$ sites, 
$n_{i\sigma}=c^{\dagger}_{i\sigma}c^{}_{i\sigma}$,
and $b^{\dagger}_{i}$ ($b^{}_{i}$) are the corresponding bosonic 
operators for  dispersionless optical phonons.

The physics of the HHM is governed by three
competing effects: The itinerancy of the electrons ($\propto t$), 
their on-site Coulomb repulsion ($\propto U$), and the 
local EP coupling ($\propto g$). 
Since the EP interaction is retarded, the phonon frequency ($\omega_0$) 
defines a further relevant energy scale. This advises us to introduce
besides the adiabaticity ratio ($\omega_0/t$)
two dimensionless coupling constants ($ u=U/4t$ and 
$\lambda=2\varepsilon_{\rm p}/2t$ or $g^2=\varepsilon_{\rm p}/\omega_0$). 

In the single-electron case, the Holstein model~\cite{Ho59a} 
has been studied extensively as a paradigmatic model 
for polaron formation (see, e.g., Ref.~\cite{FAHW_Varenna}).
Here $\varepsilon_{\rm p}$ gives the polaron binding energy. 
The Hubbard model~\cite{Hu64a}, originally designed to describe 
ferromagnetism of transition metals, more recently has  
been in use in the context of metal-insulator transition 
and high-temperature superconductivity 
as the probably most simple model to account for strong Coulomb 
correlation effects. As yet there exits almost no exact 
(analytical) results for the full HHM~(\ref{hhm}).  
At least at half-filling, however, it has become generally 
accepted that the interplay of charge, spin and 
lattice degrees of freedom gives rise to the phase diagram sketched 
in Fig.~\ref{fig_hhm_phys}. This scenario is supported by 
dynamical mean field investigations of the HMM, which become 
reliable at least in infinite spatial dimension~\cite{DMFT}. 
Besides the properties of the ground state,
the nature of the physical excitations is puzzling as well, especially
in 1D. While one expects ``normal'' electron-hole pair excitations in the  
PI phase ($U=0$), charge (spin) excitations are known to be 
massive (gapless) in the MI state of the Hubbard model ($\lambda=0$). 
Thus, varying the control parameter $u/\lambda$, 
a cross-over from standard quasi-particle behaviour  
to spin-charge separation might be observed in the more 
general 1D HHM. 

The aim of this contribution is to establish this physical picture and 
the anticipated phase diagram of the 1D HHM. 
For these purposes we employ the Lanczos exact 
diagonalisation (ED)~\cite{CW85,WRF96}, 
kernel polynomial (KPM)~\cite{SR97,WWAF05} and density-matrix renormalisation 
group (DMRG)~\cite{Wh92} methods 
(see Jeckelmann/Fehske~\cite{JF_Varenna}). 
These numerical techniques allow  
us to obtain unbiased results for all interaction 
strengths with the full quantum dynamics of phonons 
taken into account.    

\section{Luttinger-liquid Peierls-insulator transition}
\subsection{Holstein model of spinless fermions}
In a first step, let us neglect the spin degrees of freedom. 
The resulting 1D spinless Holstein model, 
$H =  - t
      \sum_{\langle i, j\rangle}\! c_{i}^\dagger 
      c_{j}^{} 
      - g \omega_0
      \sum_i (b_i^\dagger + b_i^{}) n_{i}^{}
      + \omega_0 \sum_i b_i^\dagger b_i^{}$, 
exhibits a quantum phase transition from a metallic 
to an insulating phase at half-filling 
($N_{\rm e}=N/2$)~\cite{HF83,BGL95}, where 
the critical coupling $\lambda_{\rm c}(\omega_0)\to 0$
for $\omega_0\to 0$. In the anti-adiabatic $(\omega_0\to \infty)$ strong EP
coupling regime, the model can be transformed to the 
exactly solvable XXZ (small polaron) model~\cite{HF83}, which shows 
a transition of Kosterlitz-Thouless type. Various 
variational~\cite{ZFA89}, 
renormalisation group~\cite{CB84}, 
world-line quantum Monte Carlo~\cite{HF83} or 
Green's-function Monte Carlo~\cite{MHM96} 
methods were used to determine the phase
boundary, within which significant discrepancies occur in the adiabatic
intermediate coupling regime.  More precise ED~\cite{WF98a}  
and DMRG~\cite{BMH98} techniques yields the  
phase diagram presented in Sec.~\ref{pd}. 

\subsection{Luttinger-liquid parameters and charge structure factors}
Before we discuss the metal insulator transition in the framework of
the Holstein model we will characterise the metallic and insulating phases
in themselves. According to Haldane's Luttinger liquid (LL) 
conjecture~\cite{Ha80}, 
an 1D gapless (metallic) system of interacting fermions should belong
to the Tomonaga-Luttinger universality class. Since the Holstein model
of spinless fermions is expected to be gapless at weak couplings $g$,
we try to determine the (non-universal) LL
parameters, $K_{\rho}$ (correlation exponent) and 
$u_\rho$ (charge velocity), by performing a large-scale 
DMRG finite-size scaling analysis. 
To leading order, the ground-state energy and the charge excitation gap
of a finite system with $N$ sites scales as:  
\begin{eqnarray}
 \frac{E_0(N)}{N}&=&\varepsilon_0(\infty)-
\frac{\pi}{3} \frac{u_\rho}{2}
\frac{1}{N^{2}}\,,\label{e_scaling}\\
\Delta_{\rm c}(N)&=&E^{\pm}_0(N)-E_0(N)=\pi \frac{u_\rho}{2} \frac{1}{K_\rho}
\frac{1}{N}\,.
\label{chargegap_scaling} 
\end{eqnarray}
Here $\varepsilon_0(\infty)$ denotes the energy density of the infinite system
with $N/2$ electrons, and $E^{\pm}_0(N)$ are the ground-state energies 
with $\pm 1$ fermions away from half-filling.

The LL scaling relations (\ref{e_scaling}) and 
(\ref{chargegap_scaling}) were  
derived for the pure electronic spinless fermion model 
only~\cite{Ca84}.
\begin{figure}[t]
\begin{minipage}{0.59\linewidth}
\includegraphics[width=.95\linewidth,clip]
      {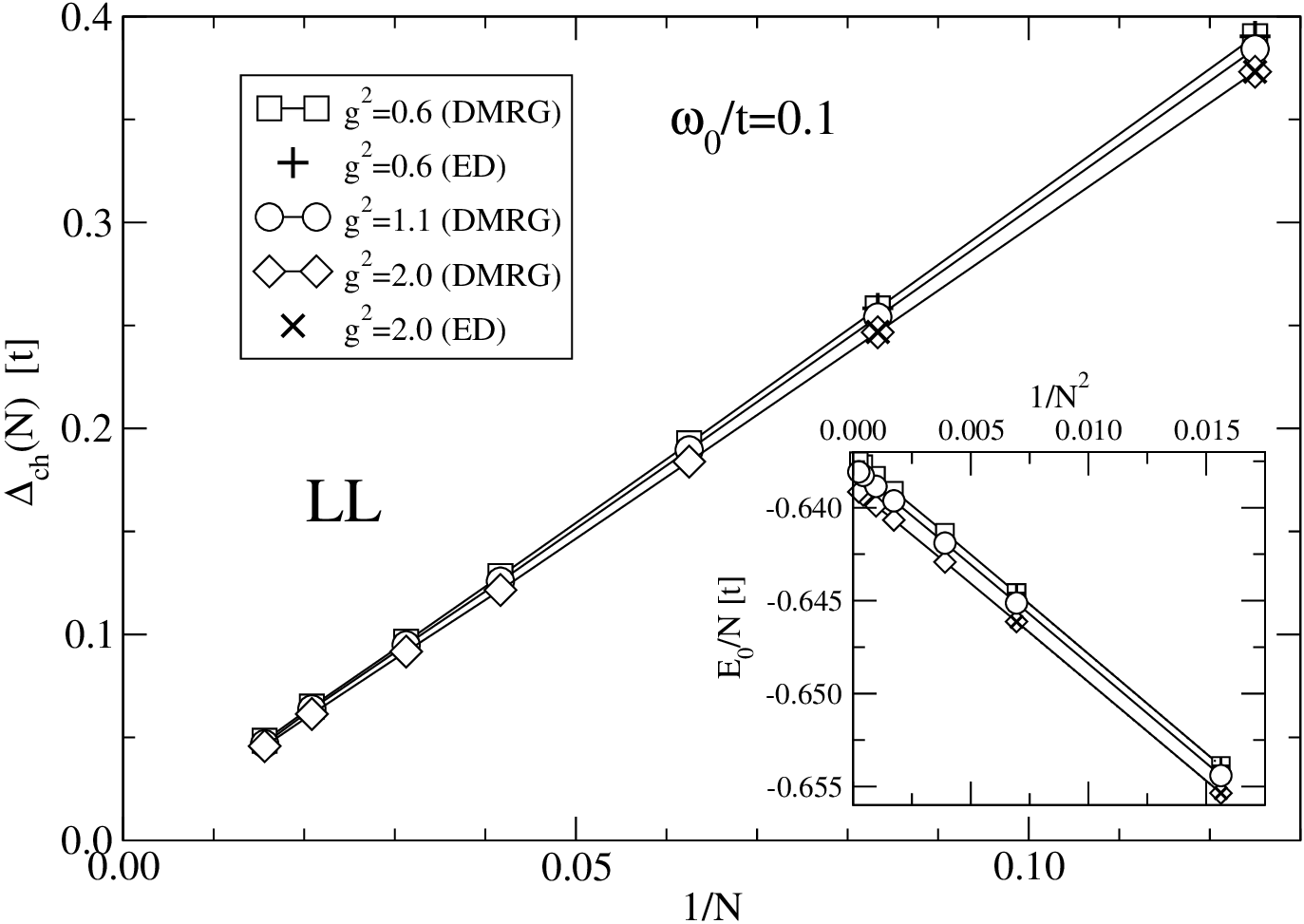}
\end{minipage} 
\begin{minipage}{0.4\linewidth}
\begin{tabular}{ccccc}\\ \hline
\rule[0mm]{0mm}{6mm}$g^2$&\multicolumn{2}{c}{
$\omega_0/t=0.1$}&
\multicolumn{2}{c}{\mbox{$\omega_0/t=10.0$}}\\
 \rule[-3mm]{0mm}{8mm}  &$K_\rho$  & 
$u_\rho/2$  & $K_\rho$    & $u_\rho/2$\\ \hline
\rule[0mm]{0mm}{6mm} 0.6   & 1.031     &  $\sim 1$  &  $\sim 1$    &  0.617\\ 
\rule[0mm]{0mm}{6mm} 2.0   & 1.055     &  0.995    &   0.949 & 0.146\\
\rule[-3mm]{0mm}{9mm} 4.0   & 1.091     &  0.963    &   0.651 & 0.028\\\hline\\
\end{tabular}
\end{minipage} 
\caption{DMRG finite-size scaling of the charge gap $\Delta_{\rm c}(N)$  
and the ground-state energy $E_0(N)$. ED 
data included for comparison. The Table gives the 
LL parameters extracted from the scaling relations. Note that $K_\rho > 1$ 
($K_\rho < 1$) in the adiabatic (anti-adiabatic) regime.}
\label{llscaling}
\end{figure}
Figure~\ref{llscaling} demonstrates, exemplarily for the adiabatic regime, 
that they also hold for the case that a finite EP coupling is included.
The resulting LL parameters are specified in the Table.
Interestingly the LL phase splits in two different regimes: 
For low phonon frequencies the effective fermion-fermion interaction 
is attractive ($K_\rho>0$), while it is repulsive  
($K_\rho<0$) for high frequencies. 
In the latter region the kinetic energy ($\propto u_{\rho}$)
is strongly reduced and the charge carriers behave like (small) polarons.
In between, there has to be a point where the LL is made 
up of (almost) non-interacting particles ($K_\rho=1$).       
The LL scaling breaks down just at the critical coupling
$g_{\rm c}(\omega_0/t)$, signalling the transition to the CDW state. 
We found  $g_{\rm c}^2(\omega_0/t=0.1)\simeq 7.84$ 
and $g_{\rm c}^2(\omega_0/t=10)\simeq 4.41 $.

\begin{figure}[t]
\begin{minipage}{0.74\linewidth}
\includegraphics[width=\linewidth,clip]{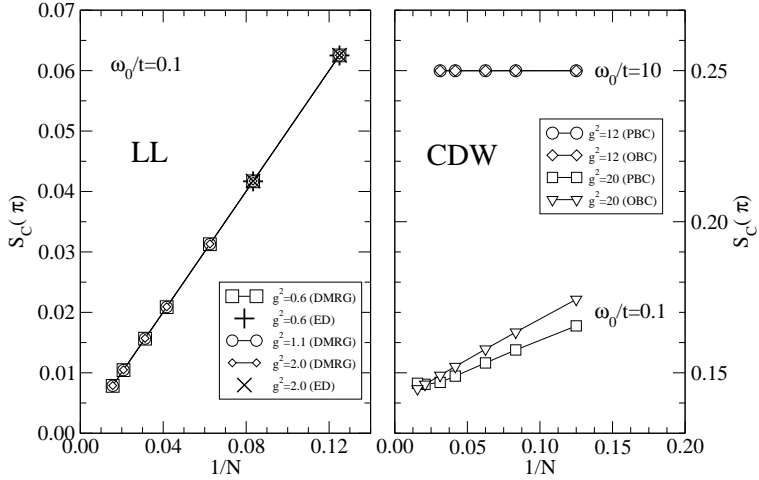}
\end{minipage}\hspace*{0.2cm} 
\begin{minipage}{0.25\linewidth}\vspace*{4cm}  
\caption{DMRG scaling of the charge structure factor $S_{\rm c}(\pi)$ using 
periodic (PBC) and open (OBC) boundary conditions.}\label{fscf} 
\end{minipage}
\end{figure}

Figure~\ref{fscf} proves the existence of CDW long-range order 
above $g_{\rm c}$. Here the staggered charge structure factor 
\begin{equation}\label{scf}
S_{\rm c}(\pi)=\frac{1}{N^2}\sum_{i,j}(-1)^j\langle (n_i-\frac{1}{2})
(n_{i+j}-\frac{1}{2})\rangle 
\end{equation}
unambiguously scales to a finite value in the thermodynamic limit
($N\to \infty$). 
In contrast we have $S_{\rm c}(\pi)\to 0$ in the metallic regime  
($g<g_{\rm c}$). Note that such a finite-size scaling, including 
dynamical phonons, is definitely out of the 
range of any ED calculation.  

\subsection{Phase diagram of the Holstein model of spinless fermions}
\label{pd}
\begin{figure}[t]
\centerline{\includegraphics[width=.8\linewidth,clip]
{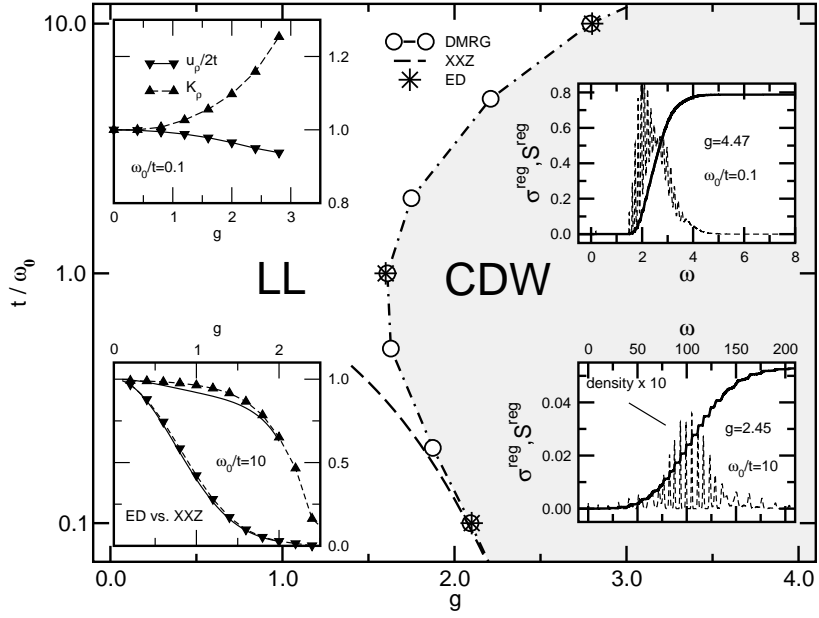}}
\caption{Ground-state phase diagram of the 1D half-filled
Holstein model of spinless fermions,
showing the boundary between the LL  
and CDW states obtained by ED and DMRG approaches. 
The dashed line gives the asymptotic
result for the XXZ model. Left insets show the 
LL parameters  $u_{\rho}$ and $K_{\rho}$ as a function of the 
EP coupling $g$ in the metallic regime; right
insets display for a six-site chain the regular part of the optical 
conductivity $\sigma^{\rm reg}(\omega)$ (dotted lines) and the integrated spectral 
weight ${\cal S}^{\rm reg}(\omega)=\int_0^{\omega}d\omega^{\prime}
\sigma^{\rm reg}(\omega^{\prime})$ (solid lines) in the CDW region.}\label{f_pd} 
\end{figure}

Figure~\ref{f_pd} gives the position of the 
LL-CDW phase boundary in the $g$--$\omega_0^{-1}$ plane.  
For weak EP interactions the system is a metal
with LL parameters depending on both the coupling strength
and the phonon frequency. Increasing $\omega_0$ at fixed $g$ (which
nonetheless means that the EP coupling parameter $\lambda$ becomes larger) 
the cross-over from an attractive LL (adiabatic regime) to a 
repulsive LL (anti-adiabatic regime) takes place at about 
$\omega_0/t \simeq 1$. Keeping $\omega_0$ fixed we enter the
CDW phase at a critical EP coupling $g_{\rm c}(\omega_0)$.
ED, cluster perturbation theory~\cite{HWBAF05} and projector-based 
renormalisation methods~\cite{SHBWF05} 
reveals the softening of the optical phonon 
at the Brillouin-zone boundary, at least in the adiabatic regime, 
which can be understood as precursor effect of the gap 
formation. Note that in $D=\infty$, the opening of the electronic gap is 
accompanied by the appearance of a low-energy phonon peak in 
the total phononic spectral function~\cite{MHB02}.
  
The CDW for strong EP coupling is connected to a Peierls distortion of the 
lattice, and can be classified as traditional band insulator 
and polaronic superlattice in the strong-EP coupling adiabatic
($\omega_0/t\ll 1$)
and anti-adiabatic ($\omega_0/t\gg 1$) regimes, respectively. 
Extremely valuable information about the CDW state
can be obtained by analysing the regular part of the optical conductivity,
\begin{equation}
\label{sigreg} 
\sigma^{\rm reg}(\omega)=\sum_{m > 0}
\frac{|\langle \psi_0^{} |\mbox{i} t \sum_{j}( c_{j}^{\dagger}
 c_{j+1}^{} - c_{j+1}^{\dagger}c_{j}^{}) |  \psi_m^{} 
       \rangle |^2}{E_m-E_0} \;\delta[\omega -(E_m-E_0)]\,,
\end{equation}
which is connected due to finite-frequency optical
transitions to excited quasi-particle states $|\Psi_m^{}\rangle$.
In Eq.~(\ref{sigreg}), $\sigma^{\rm reg}(\omega)$ is given in units 
of $\pi e^2 $ and we have omitted an $1/N$ prefactor. 
The evaluation of dynamical correlation functions
like $\sigma^{\rm reg}(\omega)$ can be carried out 
by means of the very efficient and numerically stable 
ED-KPM algorithm~\cite{WWAF05}. 
The optical absorption spectra shown in Fig.~\ref{f_pd}
elucidate the different  nature of the CDW for small and large
adiabaticity ratios. 
In the adiabatic region  the most striking feature is the sharp 
absorption threshold and large spectral weight
contained in the incoherent part of optical conductivity. 
In the anti-adiabatic regime the CDW is basically a
state of alternate self-trapped polarons, which means that 
the electrons are heavily dressed by phonons. 
Since the renormalised band dispersion is extremely narrow, 
finite-size gaps are reduced as well. Therefore, 
$\Delta_{\rm opt}$ read off from Fig.~\ref{f_pd}
correctly gives the CDW gap.

\section{Peierls-insulator Mott-insulator transition}
\subsection{Ground-state properties}
Now we include the spin degrees of freedom and ask for 
the effect of a finite Coulomb interaction. The ground state of the 
pure Holstein model ($U=0$) is a Peierls distorted state with staggered charge 
order, i.e. alternating empty and doubly occupied sites,
for  $g>g_{\rm c}(\omega_0)$~\cite{JZW99,FKSW03}.
As in the Holstein model of spinless fermions, quantum phonon fluctuations
destroy the Peierls state for $g<g_{\rm c}$.  
\begin{figure}[t]
\begin{minipage}{0.58\linewidth}
\includegraphics[width=.9\linewidth,clip]{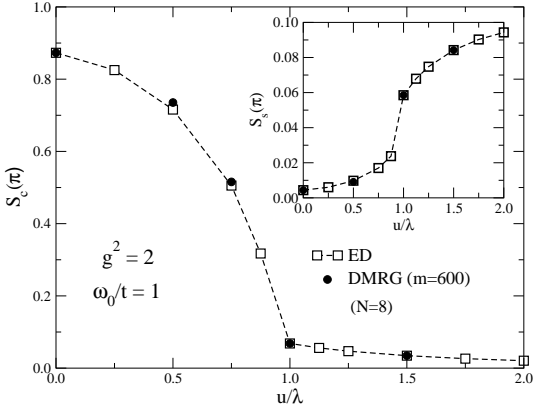}
\end{minipage}\hspace*{0.3cm} 
\begin{minipage}{0.35\linewidth}
\vspace*{2cm}  
\caption{Spin and charge 
  structure factors at $q=\pi$ in the half-filled 1D 8-site HHM
  (\ref{hhm}) with PBC for different $u$ at $\lambda=1$. 
Squares denote ED results, filled circles show DMRG calculations
  with $m=600$ target states and six pseudo-sites.}\label{scss}
\end{minipage} 
\end{figure}
\begin{figure}[b]
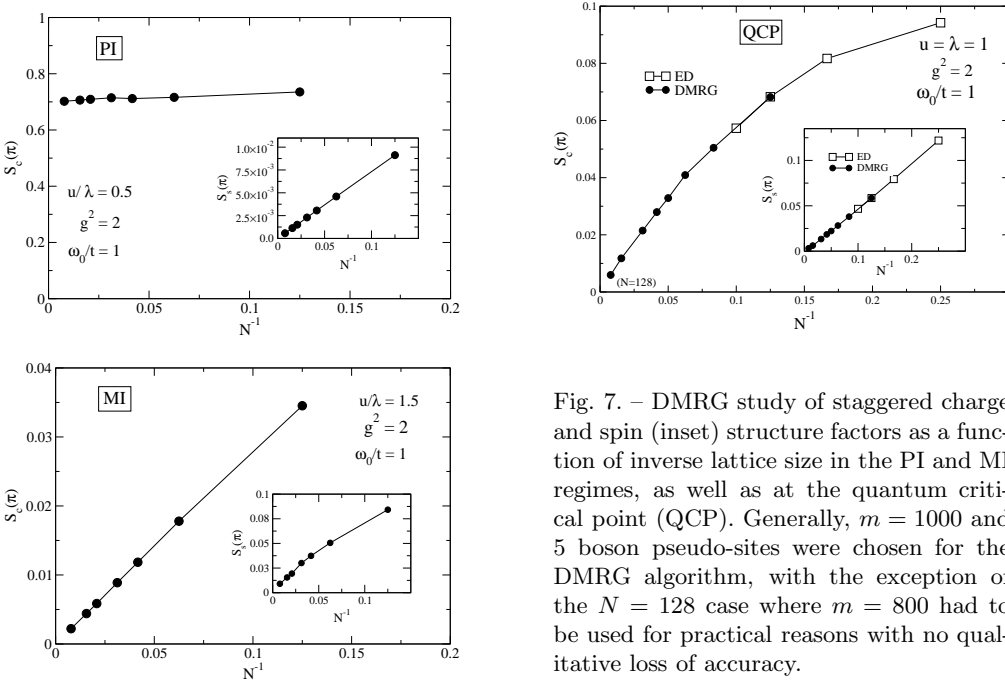

\begin{minipage}{0.45\linewidth}
\includegraphics[width=\linewidth,clip]
{fehske_jeckelmann_f7a.eps}\\[0.3cm] 
\includegraphics[width=\linewidth,clip]
{fehske_jeckelmann_f7b.eps} 
\end{minipage} 
\hspace*{1cm}
\begin{minipage}{0.45\linewidth}
\includegraphics[width=\linewidth,clip]
{fehske_jeckelmann_f7c.eps}\\[.0cm]  
\caption{DMRG study of staggered charge and  spin (inset) 
  structure factors as a function of inverse lattice size in the PI and MI 
  regimes, as well as at the quantum critical point (QCP).  
  Generally, $m=1000$   
  and $5$ boson pseudo-sites were chosen for the DMRG algorithm, with
  the exception of the $N=128$ case where $m=800$ 
  had to be used for practical reasons with no qualitative loss of accuracy.}
\label{scssscaling}
\end{minipage} 
\end{figure}

The charge structure factor, 
$S_\mathrm{c}(\pi)$ [cf. Eq.~(\ref{scf}], and spin structure factor, 
\begin{equation}
S_\mathrm{s}(\pi) = \frac{1}{N^2}\sum_{i,j} 
(-1)^j\langle S_i^zS_{i+j}^z\rangle \;\;\mbox{with}\;\;\; 
S_i^z={1\over 2}(n_{i\uparrow}-n_{i\downarrow})\,,
\label{scs}
\end{equation}
shown in  Fig.~\ref{scss} for the full HHM, 
indicate pronounced CDW and weak SDW correlations 
provided $u/\lambda < 1$. Increasing the Hubbard interaction
$u$ at fixed EP coupling $\lambda$ and frequency
$\omega_0$, the CDW correlations become strongly suppressed,
whereas the spin structure factor at $q=\pi$ is enhanced.

The results depicted in Fig.~\ref{scss} are obtained for a rather
small eight-site chain with PBC. In order to conclude about a  
possible existence of charge and/or spin long-range order
we have calculated $S_{\rm c}(\pi)$ and
$S_{\rm s}(\pi)$ for different system sizes
and performed a finite-size scaling of our DMRG results 
(see Fig.\ref{scssscaling}). 
In the PI phase, $S_{\rm c}(\pi)$ shows almost no dependence on
the size of the system, indicating true CDW long-range order, whereas
$S_{\rm s}(\pi)$ obviously scales to zero as $N\to\infty$.  
By contrast, in the MI regime our data provides strong evidence 
for vanishing charge but also spin order in the thermodynamic limit.
Clearly the MI is characterised by short-ranged antiferromagnetic
spin correlations but nevertheless the staggered spin-spin correlation
function shows a slow (algebraic) decay at large distances.

\begin{figure}[t]
\begin{minipage}{0.6\linewidth}
\includegraphics[width=\linewidth,clip]
{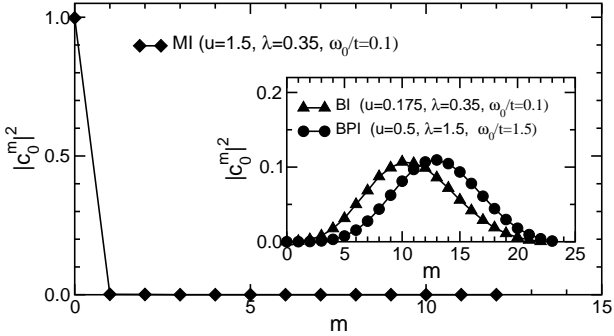}
\end{minipage}\hspace*{0.5cm}
\begin{minipage}{0.38\linewidth}\vspace*{1cm}
\caption{Phonon distribution, $|c_0^m|^2$, for typical model parameters 
characterising the MI, band insulator (BI) and bipolaronic insulator(BPI)
Peierls phases (inset). 
Note that we have separated the $Q=0$ centre of mass phonon.}
\label{phondis}
\end{minipage}
\end{figure}
Figure~\ref{phondis} gives the phonon distribution function, i.e.
the weights of the $m$-phonon state for different ground states
of the HHM. First of all the results demonstrate that our
Hilbert space optimisation and truncation procedure is well-controlled
in the sense that states with larger number of phonons, as accounted
for in the calculations, have negligible spectral weight. Of course,
the number of phonons which have to be taken into account depends on 
the physical situation. Whereas the ground state of the MI is
basically a zero-phonon state, multi-phonon states become increasingly 
important if $u/\lambda$ is reduced, i.e. in the PI state.

\subsection{Optical response}
\begin{figure}[t]
\centerline{\includegraphics[width=.7\linewidth,clip]
{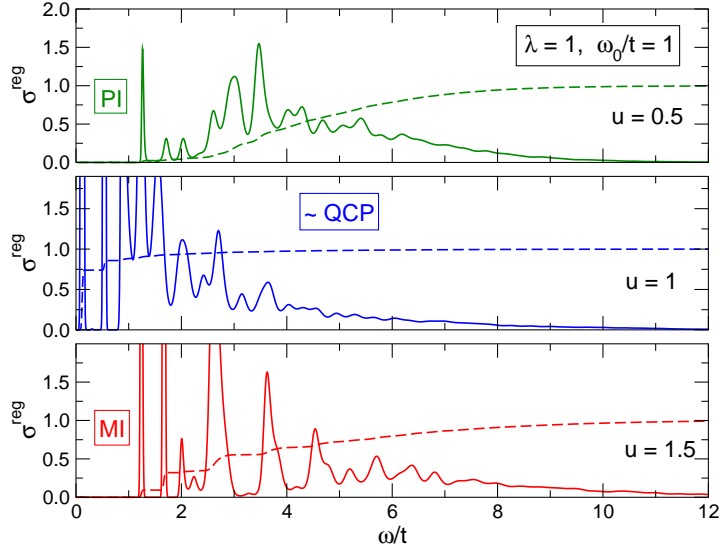}}
\caption{Optical conductivity in the Holstein Hubbard model (ED, $N=8$, PBC).
Dashed lines give the integrated spectral weight normalised 
by $S^{\rm reg}(\infty)$. $S^{\rm reg}(\omega)/S^{\rm  reg}(\infty)$ 
is a natural measure for  
the relative weight of the different optical absorption processes. 
Note that  the critical coupling 
is identical to the point where $S_{\rm c}(\pi)$ sharply drops 
(cf. Fig.~\ref{scss}).}
\label{optlei}
\end{figure}
The evolution of the optical conductivity  $\sigma^{\rm reg}(\omega)$
going from the PI to the MI phase by increasing $u/\lambda$ is illustrated in
Fig.~\ref{optlei}. In the PI regime the electronic excitations are gapped
due to the CDW formation. Excitonic gap states may occur 
in the process of structural relaxation. Because the PI ground state 
is a multi-phonon state, we find a gradual rise of the integrated 
spectral weight of $\sigma^{\rm reg}(\omega)$.

At the QCP the optical gap $\Delta_{\rm opt}$ closes.
Since we are in the non-adiabatic regime ($\omega_0\simeq t$), 
the lowest optical excitations have mainly pure 
electronic character in the vicinity of $(u/\lambda)_{\rm c}$,
i.e. the gap is closed by a state having large electronic spectral 
weight. Due to the selection rules for optical transitions the 
PI-MI transition necessarily implies a ground-state level crossing 
with a site-parity change. The site inversion symmetry operator $P$ is defined
by $Pc_{i\sigma}^\dagger P^\dagger=c_{N-i\sigma}^\dagger$ ($N=4n$). 
We have explicitly verified that the parity is $P=+1$  ($P=-1$)
in the PI (MI) phase.

In the MI the optical gap is by its nature a correlation gap. The lower 
panel in Fig.~\ref{optlei}  shows clearly that $\sigma^{\rm reg}(\omega)$ 
is dominated by excitations which can be 
related to those of the pure Hubbard model. 
In addition, phononic sidebands with low spectral
weight appear.

\subsection{Photoemission spectra}
Next we study the spectral density 
of single-particle excitations associated with 
the injection of a spin-$\sigma$ electron with wave number 
$K$, $A_{K\sigma}^{+}(\omega)$ (inverse photoemission (IPE)),
and the corresponding quantity 
for the emission of an electron, $A_{K\sigma}^{-}(\omega)$ 
(photoemission (PE)), where  
\begin{equation}
A_{K\sigma}^{\pm}(\omega) =  \sum_{m} 
|\langle \psi_m^{\pm}|c_{K\sigma}^{\pm} 
|\psi_0^{}\rangle|^2 \,\delta [\,\omega\mp(E_m^{\pm}-E_0^{})] 
\label{ipe}
\end{equation}
with $c_{K\sigma}^{+}=c_{K\sigma}^{\dagger}$  
and $c_{K\sigma}^{-}=c_{K\sigma}^{}$. 
$|\psi_0^{}\rangle$, $E_0^{}$ refer to the ground state of the system 
with $N_{\rm e}=N$ electrons and $|\psi_m^{\pm}\rangle$ ($E_m^{\pm}$)
are eigenstates (energies) of the $(N_{\rm e} \pm 1)$-particle system. 
Adding the spectral densities 
of (photo-) emission and absorption 
we obtain the spectral function 
$A_{K\sigma}(\omega)=A_{K\sigma}^+(\omega)+ A_{K\sigma}^-(\omega)$,
which obeys various sum rules and allows for a connection to
angle-resolved photoemission spectroscopy (ARPES).

\begin{figure}[t]
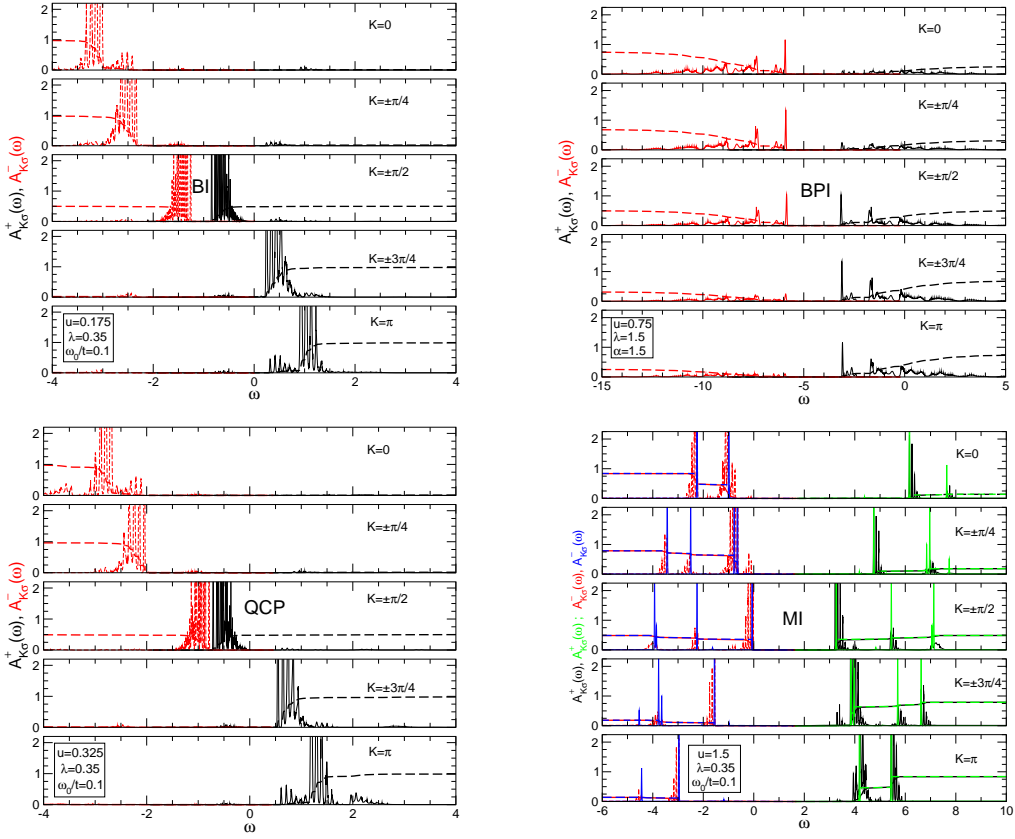

\begin{minipage}{0.45\linewidth}
\includegraphics[width=\linewidth,clip]
{fehske_jeckelmann_f10a.eps}\\[0.3cm] 
\includegraphics[width=\linewidth,clip]
{fehske_jeckelmann_f10b.eps} 
\end{minipage} 
\hspace*{1cm}
\begin{minipage}{0.45\linewidth}
\includegraphics[width=\linewidth,clip]
{fehske_jeckelmann_f10c.eps}\\[0.3cm]
\hspace*{.2cm}\includegraphics[width=.975\linewidth,clip]
{fehske_jeckelmann_f10d.eps}
\end{minipage}
\caption
{Wave-number-resolved spectral densities  
for photoemission ($A_{K\sigma}^-(\omega)$; dashed (red) lines) 
and inverse photoemission ($A_{K\sigma}^+(\omega)$; solid (black) lines)
obtained by applying our ED-KPM scheme to the HHM.  
Shown are typical results obtained  
for the case of a Peierls band insulator (BI) at $\omega_0/t\ll 1$ (left) 
and bipolaronic insulator (BPI) at $\omega_0/t\gg 1$ (right). 
The corresponding integrated densities
$S_{K\sigma}^{\pm}(\omega)$ are given by dashed lines.
The lower panels give the PE and IPE spectra 
near the PI-MI transition point ($u/\simeq 1$) and 
in the Mott insulating state ($u/\lambda\gg 1$).  
Data for the pure Hubbard model (blue and green lines)
were shifted by $-(\varepsilon_{\rm p} N^2_{\rm e}/N)$ and included  
for comparison. Results are taken from Ref.~\cite{FWHWB04}.} 
\label{ipespectra}
\end{figure}

Figure~\ref{ipespectra} displays the IPE and PE spectra for the 
HHM at the allowed wave numbers of our finite system:
$K=0,\, \pm\pi/4,\,\pm\pi/2, \pm 3\pi/4$, 
and $\pi$. To reliably monitor a band splitting 
induced by the Hubbard and EP couplings it is necessary to guarantee that 
the Fermi momenta $K_F=\pm\pi/2$ are occupied, which is  
the case for $N=4l$ ($l$ integer, PBC).

The most prominent feature we observe in the PI regime is 
the opening of a gap at $K=\pm \pi/2$. 
For the BI a rather broad (I)PE 
signatures appear.  Within these excitation bands the spectral 
weight is almost uniformly distributed, 
which is a clear signature of the multi-phonon
absorption and emission processes that accompany every
single-particle excitations in the PI. 
The lineshape reflects the (Poisson-like) distribution 
of the phonons in the ground state. The lower and upper 
band closely follow a (slightly renormalised) cosine dispersion. 
The situation changes radically if the insulating behaviour
is associated with localised bipolarons forming a CDW state 
(see Fig.~\ref{ipespectra}, upper right panel). Due to strong polaronic effects
an almost flat band dispersion results with exponentially  
small (electronic) quasi-particle weight. Now the dominant 
peaks in the incoherent part of the (I)PE spectra
are related to multiples of the (large) bare phonon
frequency. 

If we increase the Hubbard interaction at fixed EP coupling 
strength the Peierls gap weakens and  finally  closes at about 
$(u/\lambda)_c\simeq 1$, which marks the PI-MI cross-over. 
This is the situation shown in the lower left panel. 
Approaching the QCP, 
the ground state and the first excited state become degenerate.
The QCP is characterised by gapless charge
excitations at the Fermi momenta but perhaps should not be considered
as metallic because the Drude weight in the case of a degenerate
ground state is ill-defined~\cite{Ko64}. 

If the Hubbard interaction is further increased, i.e. the Coulomb 
repulsion overcomes the attractive on-site 
EP coupling the electronic band structure becomes gapped again 
and a MI state develops (see lower right panel). 
The Mott-Hubbard correlation gap almost 
coincides with the optical gap $\Delta_{\rm opt}$ determined by  
evaluating the regular part of the optical conductivity
for the same parameters. The form of the spectra, however, 
is quite different from PI case. Contrary to the BI phase 
in the MI regime the lowest peak 
in each $K$ sector is clearly the dominant one. 
Then the dispersion of the lower (upper) Hubbard band
can be derived tracing the uppermost (lowest) excitations 
in each $K$ sector. 
Due to the finiteness 
of our system and the rather moderate value $u=1.5$,
PE (IPE) excitations with $K=\pm 3\pi/4$ and $\pi$
($K=\pm \pi/4$ and $0$) have still finite spectral weight. 
Since the spectral weight of the PE excitations with $K>\pi/2$
is expected to vanish as $N$ goes to infinity for $u\gg 1$, 
the lower Hubbard band will be completely filled, and consequently
the system behaves as an insulator at $T=0$.  
As a result of the coupling to the phonon system the electronic 
levels in each $K$ sector split,  creating phonon 
side bands. The distinct peaks are separated by
multiples of the bare phonon frequency and 
can be assigned to relaxation processes of the $Q=0$ phonon
modes~\cite{SHBWF05}.  The number of phonons involved is controlled by $g^2$.  
$S_{K\sigma}^{\pm}(\omega)$ shows clearly that 
the total spectral weight of the resulting excitation
bands equals the weight of the respective electronic 
excitations in the pure Hubbard model.
Interestingly, mediated by $Q\neq 0$ phonons,
there appear ``shadows'' of the bands composed of a 
dominant electronic excitation and phonon satellites
in a certain $K$ sector (e.g. $K=\pm \pi/2$) 
in the other $K$ sectors (e.g. $K=\pm \pi/4$).  
These signatures give rise to a weak  ``breather-like'' 
excitation being almost dispersionless in the Brillouin zone. 
The formation of  
quantum breathers was proposed by W. Z. Wang {\it et al.}~\cite{WBGS98},
but without doubt has not been detected so far. 

\begin{figure}[t]
\centerline{\includegraphics[width=.7\linewidth,clip]
{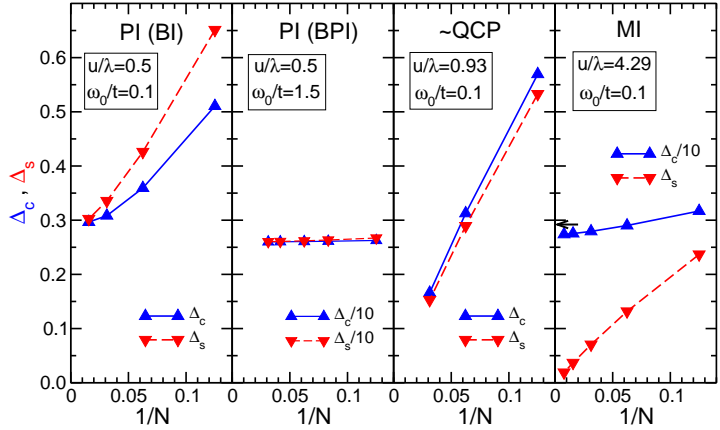}}
\caption{DMRG finite-size scaling of spin- and charge excitation
     gaps in the HHM at $\lambda=0.35$ and $\omega_0/t=0.1$,
where open boundary conditions were used. 
The accessible system sizes are smaller at
larger $\lambda/u$, where an increasing number 
of (phononic) pseudo-sites is required to reach  
convergence with respect to the phonons.  
The arrow marks the value of the optical gap $\Delta_{\rm opt}$
for the Bethe ansatz solvable 1D Hubbard model, 
which is given  by $\Delta_{\rm opt}/4t = u - 1 + \ln(2)/2u$
in the limit of large $u>1$~\cite{Ov70}.} 
\label{cgsg}
\end{figure}

\subsection{Many-body excitation gaps}
Since many-body gaps to excited states 
form the basis for making contact with experimentally measurable 
excitation gaps and can also be used to characterise
different phases of the HHM, we finally determine the  
charge and spin gaps,
 \begin{eqnarray}
 \label{cg}
 \Delta_{\rm c}&=&E_0^{+}(1/2)+E_0^{-}(-1/2)-2E_0(0)\\
 \label{sg}
 \Delta_{\rm s}&=&E_0(1)-E_0(0)\,,
 \end{eqnarray}
using DMRG. 
Here $E_0^{(\pm)}(S^z)$ is the ground-state 
energy of the HHM at half-filling (with $N_{\rm e}=N\pm 1$) 
particles in the sector with total spin-$z$ component $S^z$.

The scaling of $\Delta_{\rm c}$ and $\Delta_{\rm s}$ is shown in
fig.~\ref{cgsg}.   
Obviously, $\Delta_{\rm c}$ and $\Delta_{\rm s}$ are finite
in the PI and will converge further to the same value as $N\to \infty$. 
Both gaps seem to vanish at the QCP 
of the HHM with finite-frequency phonons, but in the critical
region the finite-size scaling is extremely delicate.
In the MI we found a finite charge excitation gap, which
in the limit $u/\lambda\gg 1$ scales to the optical gap 
of the Hubbard model, whereas the extrapolated spin gap 
remains zero. This can be taken as a clear indication for
spin charge separation.
\section{Summary}
In this report we have addressed the important problem of quantum phase 
transitions in one-dimensional strongly coupled electron-phonon systems.
As a generic model we analysed the Holstein Hubbard model at half-filling. 
Applying numerical diagonalisation methods we obtained, 
by the use of present-day leading-edge supercomputers, 
basically exact results for 
both ground-state and spectral properties in the overall 
region of electron-electron/electron-phonon
coupling strengths and phonon frequencies.

For the spinless Holstein model we found that for weak EP 
couplings the system resides in a metallic (gapless) phase described by two 
non-universal Luttinger-liquid parameters. The renormalised 
charge velocity and the correlation exponent are obtained 
by DMRG from finite-size scaling relations, fulfilled with great accuracy. 
The Luttinger liquid phase splits in an attractive and repulsive 
regime at low and high phonon frequencies, respectively. 
Here the polaronic metal, realised for repulsive interactions, 
is characterised by a strongly reduced mobility of the charge carriers.    
Increasing the EP interaction, a cross-over
between Luttinger-liquid and charge-density-wave behaviour 
is found. The transition to the CDW state is accompanied 
by significant changes in the optical response of the system. 
Most notably seems to be the substantial spectral weight transfer from the 
Drude to the regular part of the optical conductivity, 
indicating the increasing importance of inelastic scattering processes 
in the CDW (PI) regime.

\begin{figure}[h]
\begin{minipage}{0.45\linewidth}
\includegraphics[width=\linewidth,clip]{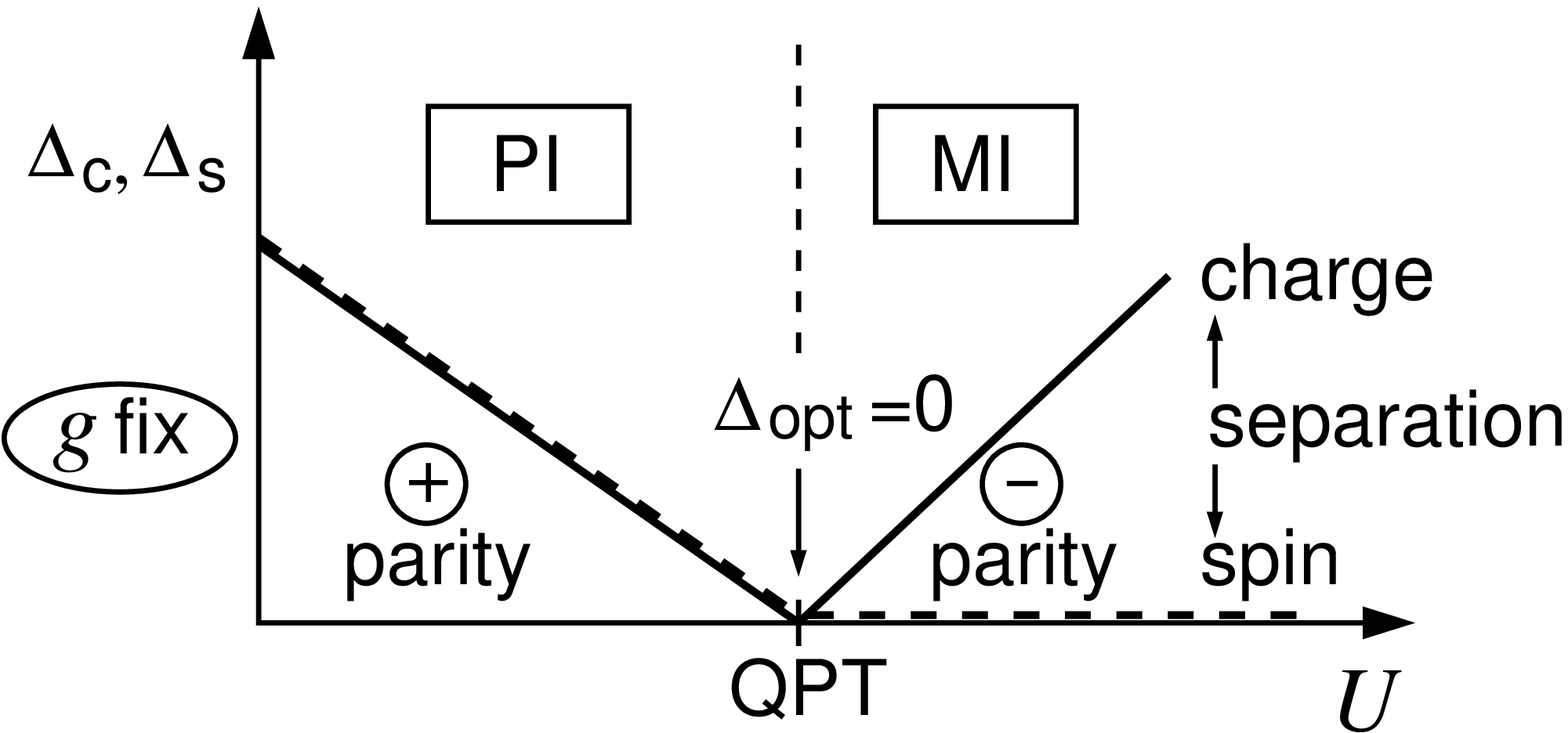}
\end{minipage}\hspace*{1cm} 
\begin{minipage}{0.45\linewidth}  
\includegraphics[width=\linewidth,clip]{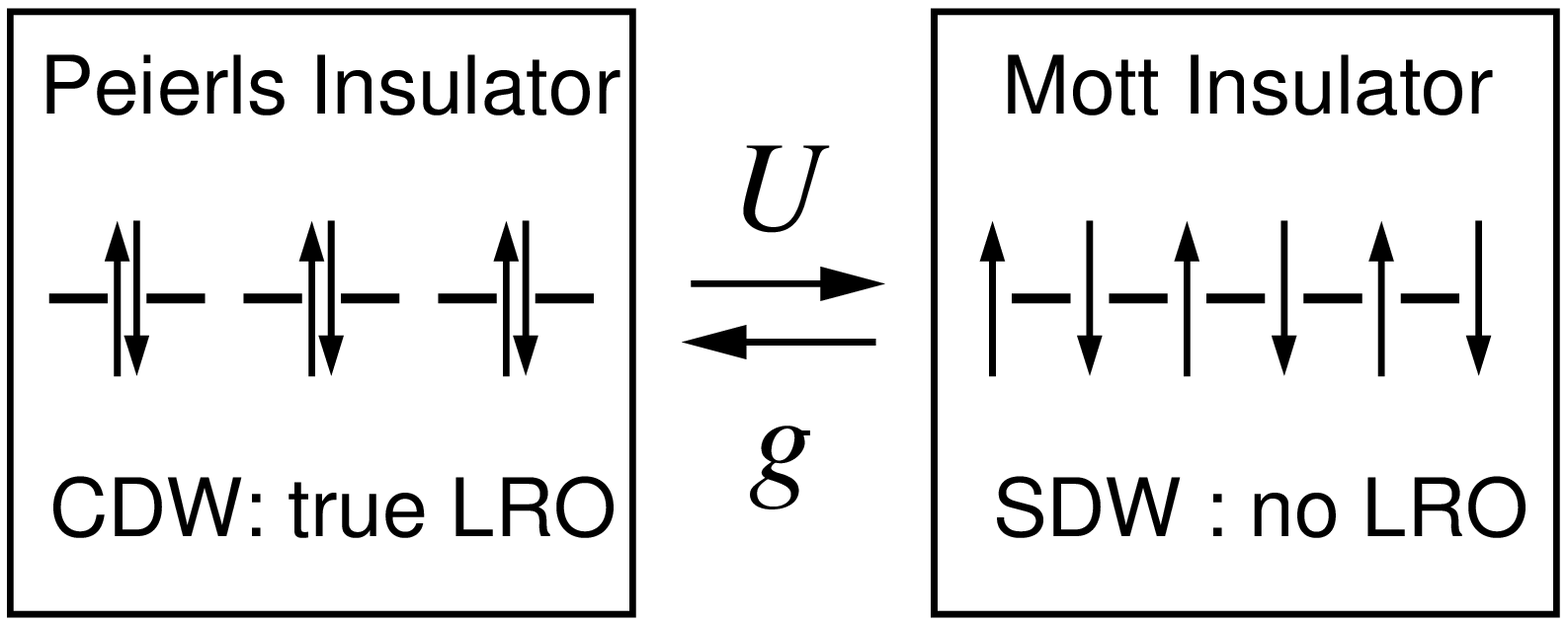}
\end{minipage}  
\caption{Sketch of the PI-MI quantum phase transition in the 
HHM.}
\label{pp}
\end{figure}

For the much more involved Holstein Hubbard model,
with respect to the metal the electron-electron interaction
favours a Mott insulating state whereas the EP coupling
 is responsible for the Peierls insulator to occur
(see Fig.~\ref{pp}). True long-range (CDW) order is established
in the PI phase only. The PI typifies a band insulator in the adiabatic 
weak-to-intermediate coupling range or a bipolaronic
insulator for non-to-antiadiabatic strong-coupling.
Our results for the single-particle spectra indicate 
that while polaronic features emerge only at strong EP 
couplings, pronounced phonon signatures, such as multi-phonon bound   
states inside the CDW gap, can be found in the Mott insulating regime as well. 
This might be of great importance for interpreting 
photoemission experiments of low-dimensional materials 
such as MX-chain compounds. 
The optical conductivity shows different absorption features
in the MI and PI as well and signals that the quantum phase transition
between these phases is connected to a change in the ground-state
site-parity eigenvalue (of a finite HHM systems with PBC).
From our conductivity data we found evidence for only one critical point
separating Peierls and Mott insulating phases in the Holstein Hubbard model
with dynamical phonons. This differs from the results obtained in 
the adiabatic limit ($\omega_0=0$), where two successive transitions 
have been detected for weak couplings $u$, $\lambda \ll 1$~\cite{FKSW03}.  
The Peierls to Mott transition scenario is corroborated
by the behaviour of the spin- and charge excitation gaps.
From a DMRG finite-size scaling we found  
$\Delta_{\rm c}=\Delta_{\rm s}$ and $\Delta_{\rm c}>
\Delta_{\rm s}=0$ in the PI and MI, respectively.  
The emerging physical picture can be summarised 
by the  phase diagram shown in Fig.~\ref{pp}.
\acknowledgments
We would like to thank A. Alvermann, K. W. Becker, A. R. Bishop, 
H. B\"uttner, F. Essler, F. Gebhard, 
F. G\"ohmann, G. Hager, M. Hohenadler, A. P. Kampf, J. Loos, M. Sekania,
S. Sykora, A. Wei{\ss}e, G. Wellein, and S. R. White 
for valuable discussions.

\begin{thebibliography}{10}
\providecommand{\url}[1]{\texttt{#1}}
\providecommand{\urlprefix}{URL }

\bibitem{TNYS90}
\textsc{Tsuda N.}, \textsc{Nasu K.}, \textsc{Yanese A.} and \textsc{Siratori
  K.}, \emph{Electronic Conduction in Oxides} (Springer-Verlag, Berlin), 1990.

\bibitem{Pe55}
\textsc{Peierls R.}, \emph{Quantum theory of solids} (Oxford University Press,
  Oxford), 1955.

\bibitem{Ho59a}
\textsc{Holstein T.}, \emph{Ann. Phys. (N.Y.)}, \textbf{8} (1959) 325;
\textbf{8} (1959) 343.

\bibitem{FAHW_Varenna}
\textsc{Fehske H.}, \textsc{Alvermann A.}, \textsc{Hohenadler
    M.} and \textsc{Wellein G.}, 
Proc. Int. School of Physics ``Enrico Fermi'', Course CLXI,
{\sl Polarons in Bulk Materials and Systems with Reduced Dimensionality},
Eds. G. Iadonisi, J. Ranninger, G. de Filippis,  (IOS Press, 
Amsterdam, Oxford, Tokio, Washington DC) 2006, pp 285-296.


\bibitem{Hu64a}
\textsc{Hubbard J.}, \emph{Proc. Roy. Soc. London, Ser. A}, \textbf{277} (1964)
  237.

\bibitem{DMFT}
\textsc{Capone M.} and \textsc{Cuichi S.}, 
\emph{Phys. Rev. Lett.}, \textbf{91} (2003) 186405;
\textsc{Sangiovanni G.}, \textsc{Capone M.}, 
\textsc{Castellani C.} and \textsc{Crilli M.}, 
\emph{Phys. Rev. Lett.}, \textbf{94} (2005) 026401.



\bibitem{CW85}
\textsc{Cullum J.K.} and \textsc{Willoughby R.A.}, \emph{Lanczos Algorithms for
  Large Symmetric Eigenvalue Computations}, volume I \& II, (Birkh\"auser,
  Boston), 1985.

\bibitem{WRF96}
\textsc{Wellein G.}, \textsc{R\"oder H.} and \textsc{Fehske H.}, \emph{Phys.
  Rev. B}, \textbf{53} (1996) 9666.

\bibitem{SR97}
\textsc{Silver R.N.} and \textsc{R\"oder H.}, \emph{Phys. Rev. E}, \textbf{56}
  (1997) 4822.

\bibitem{WWAF05}
\textsc{B{\"a}uml B.}, \textsc{Wellein G.} and \textsc{Fehske H.}, \emph{Phys.
  Rev. B}, \textbf{58} (1998) 3663; 
\textsc{Wei{\ss}e A.}, \textsc{Wellein G.}, \textsc{Alvermann A.} and
  \textsc{Fehske H.}, \emph{Rev. Mod. Phys.}, \textbf{78} (2006) 275.

\bibitem{Wh92}
\textsc{White S.R.}, \emph{Phys. Rev. Lett.}, \textbf{69} (1992) 2863; 
\textsc{Hager G.}, \textsc{Jeckelmann E.}, \textsc{Fehske H.}, \textsc{Wellein
  G.} and \textsc{Hager G.}, \emph{J. of Comp. Phys.}, \textbf{194} (2004) 795.

\bibitem{JF_Varenna}
\textsc{Jeckelmann E.} and \textsc{Fehske H.}, Proc. Int. School of Physics ``Enrico Fermi'', Course CLXI,
{\sl Polarons in Bulk Materials and Systems with Reduced Dimensionality},
Eds. G. Iadonisi, J. Ranninger, G. de Filippis,  (IOS Press, 
Amsterdam, Oxford, Tokio, Washington DC) 2006, pp 247-284.

\bibitem{HF83}
\textsc{Hirsch J.E.} and \textsc{Fradkin E.}, \emph{Phys. Rev. B}, \textbf{27}
  (1983) 4302.

\bibitem{BGL95}
\textsc{Benfatto G.}, \textsc{Gallavotti G.} and \textsc{Lebowitz J.L.},
  \emph{Helv. Phys. Acta}, \textbf{68} (1995) 312.

\bibitem{ZFA89}
\textsc{Zheng H.}, \textsc{Feinberg D.} and \textsc{Avignon M.}, \emph{Phys.
  Rev. B}, \textbf{39} (1989) 9405.

\bibitem{CB84}
\textsc{Caron L.G.} and \textsc{Bourbonnais C.}, \emph{Phys. Rev. B},
  \textbf{29} (1984) 4230.

\bibitem{MHM96}
\textsc{McKenzie R.H.}, \textsc{Hamer C.J.} and \textsc{Murray D.W.},
  \emph{Phys. Rev. B}, \textbf{53} (1996) 9676.

\bibitem{WF98a}
\textsc{Wei{\ss}e A.} and \textsc{Fehske H.}, \emph{Phys. Rev. B}, \textbf{58}
  (1998) 6208.

\bibitem{BMH98}
\textsc{Bursill R.J.}, \textsc{McKenzie R.H.} and \textsc{Hamer C.J.},
  \emph{Phys. Rev. Lett.}, \textbf{80} (1998) 5607; 
\textsc{Fehske H.}, \textsc{Wellein G.}, \textsc{Hager G.}, \textsc{Wei{\ss}e
  A.}, \textsc{Becker K.W.} and \textsc{Bishop A.R.}, \emph{Physica B},
  \textbf{359-361} (2005) 699.

\bibitem{Ha80}
\textsc{Haldane F.D.M.}, \emph{Phys. Rev. Lett.}, \textbf{45} (1980) 1358.

\bibitem{Ca84}
\textsc{Cardy J.L.}, \emph{J. Phys. A}, \textbf{17} (1984) L385;
\textsc{Voit J.}, \emph{Rep. Prog. Phys.}, \textbf{58} (1997) 977.

\bibitem{HWBAF05}
\textsc{Hohenadler M.}, \textsc{Wellein G.}, \textsc{Bishop A.R.},
  \textsc{Alvermann A.} and \textsc{Fehske H.}, 
\emph{Phys. Rev. B}, \textbf{xx} (2006) xxxx.

\bibitem{SHBWF05}
\textsc{Sykora S.}, \textsc{H{\"u}bsch A.}, \textsc{Becker K.W.},
  \textsc{Wellein G.} and \textsc{Fehske H.}, \emph{Phys. Rev. B}, \textbf{71}
  (2005) 045112.


\bibitem{MHB02}
\textsc{Meyer D.}, \textsc{Hewson A.C.} and \textsc{Bulla R.}, \emph{Phys. Rev.
  Lett.}, \textbf{89} (2002) 196401; Hewson A.C.,  
Proc. Int. School of Physics ``Enrico Fermi'', Course CLXI,
{\sl Polarons in Bulk Materials and Systems with Reduced Dimensionality},
Eds. G. Iadonisi, J. Ranninger, G. de Filippis,  (IOS Press, 
Amsterdam, Oxford, Tokio, Washington DC) 2006, pp 155-175.

\bibitem{JZW99}
\textsc{Jeckelmann E.}, \textsc{Zhang C.} and \textsc{White S.R.}, \emph{Phys.
  Rev. B}, \textbf{60} (1999) 7950. 

\bibitem{FKSW03}
\textsc{Fehske H.}, \textsc{Kampf A.P.},   \textsc{Sekania M.} and
  \textsc{Wellein G.}, \emph{Eur. Phys. J. B}, \textbf{31} (2003) 11.

\bibitem{Ko64}
\textsc{Kohn W.}, \emph{Physical Review}, \textbf{133} (1964) A171.

\bibitem{FWHWB04}
\textsc{Fehske H.}, \textsc{Wellein G.}, \textsc{Hager G.}, \textsc{Wei{\ss}e
  A.} and \textsc{Bishop A.R.}, \emph{Phys. Rev. B}, \textbf{69} (2004) 165115.


\bibitem{WBGS98}
\textsc{Wang W.Z.}, \textsc{Bishop A.R.}, \textsc{Gammel J.T.} and
  \textsc{Silver R.N.}, \emph{Phys. Rev. Lett.}, \textbf{80} (1998) 3284.

\bibitem{Ov70}
\textsc{Ovchinnikov A.A.}, \emph{Sov. Phys. JETP}, \textbf{30} (1970) 1160.



\end{thebibliography}

\end{document}